\documentclass[prd,aps,secnumarabic,superscriptaddress,floatfix]{revtex4}
\usepackage{dcolumn}
\usepackage{amsmath}
\def\sc{\mbox{\rule[-5pt]{0pt}{16pt}}}

\usepackage[dvips]{graphicx,color}
\def\sb{\mbox{\rule{0pt}{11pt}}}

\def\sw{\mbox{\rule{24pt}{0pt}}}
\def\al{\alpha}
\def\be{\beta}
\def\ga{\gamma}
\def\de{\delta}

\def\Ga{\Gamma}
\def\Up{\Upsilon}
\def\Dot{\!\cdot\!}
\def\pr{\prime}

\begin{document}
\title{Potential model calculations and predictions for heavy quarkonium}
\author{Stanley F. Radford}
\affiliation{Department of Physics, State University of New York at Brockport, Brockport, NY 14420} \email{sradford@brockport.edu}
\author{Wayne W. Repko}
\affiliation{Department of Physics and Astronomy, Michigan State University, East Lansing, MI 48824} \email{repko@pa.msu.edu}
\date{\today}
\begin{abstract}
We investigate the spectroscopy and decays of the charmonium and upsilon systems in a potential model consisting of a relativistic kinetic energy term, a linear confining term including its scalar and vector relativistic corrections and the complete perturbative one-loop quantum chromodynamic short distance potential. The masses and wave functions of the various states are obtained using a variational technique, which allows us to compare the results for both perturbative and nonperturbative treatments of the potential. As well as comparing the mass spectra, radiative widths and leptonic widths with the available data, we include a discussion of the errors on the parameters contained in the potential, the effect of mixing on the leptonic widths, the Lorentz nature of the confining potential and the possible $c\bar{c}$ interpretation of recently discovered charmonium-like states. 
\end{abstract}
\maketitle

\section{INTRODUCTION}
The discovery of the J/$\psi$ in 1974 \cite{aab, abrams} marked the beginning of an extended interplay between experimental results and the phenomenological treatment of heavy mesons using quantum chromodynamics. The prediction and discovery of the various excited states of the charmonium system and later the upsilon system encouraged the investigation of a variety of approaches to understand the systematics of quarkonium spectra and decay modes. These include potential models \cite{egkkly,prs,schnit,egkly,grr1,grr2,grr3,gjrs,efg,elq,elq1,bg,bg1,bgs,ls}, effective field theory treatments \cite{bbp,bpsv1,bpsv2,bsv,bpsv3,bpsv4} and, more recently, lattice gauge theory calculations \cite{oka,dip,gott,joopr}. While the latter promise to provide an accurate nonperturbative description of mesons and baryons, there is still interest in perturbative and quasi-perturbative models, particularly in the investigation of the angular momentum and spin effects in heavy meson spectroscopy.

Over the past 25+ years, potential models have proven valuable in analyzing the spectra and characteristics of heavy quarkonium systems \cite{qwg}. Motivation for revisiting the potential model interpretation of the $c\bar{c}$ and $b\bar{b}$ systems at this time is provided by recent experimental results:
\begin{enumerate}
\item The discovery of several expected states in the charmonium spectrum ($\eta_c(2S)$ and $h_c(1P)$)
\item The discovery of new states ($X(3872)$, $X(3940)$, $Y(3940)$, $Y(4260)$) which could be a interpreted as above threshold charmonium levels
\item The discovery of the $1^3D_2$ state of the upsilon system
\item The determination of various $E_1$ widths for $c\bar{c}$ and $b\bar{b}$.
\end{enumerate}
Our objective here is to examine to what extent a semi-relativistic potential model which includes all $v^2/c^2$ and one-loop QCD corrections can fit the below threshold $c\bar{c}$ and $b\bar{b}$ data - both spectra and decay widths - and accommodate the new above threshold states.

One of the earliest and simplest non-relativistic potential models for heavy quarkonia is that of Eichten, {\it et al.} \cite{egkkly}, the Cornell model. This model contains a long range linear term to provide quark confinement, a short distance Coulomb-like term inspired by the zeroth-order QCD interaction and a treatment of states above the continuum threshold. In this approach, splittings within triplet state multiplets of a given orbital angular momentum $L$ are generated indirectly by differences in the couplings of states with different total angular momentum $J$ to the continuum. By including relativistic corrections to the potential corresponding to scalar or vector particle exchange \cite{prs,schnit}, it is possible to generate interaction terms depending on orbital and spin angular momenta of the quarks without introducing additional parameters. These interaction terms provide a direct mechanism for splitting the various $J$ states of a triplet angular momentum multiplet.  

Further refinement of the potential requires a consideration of how the  orbital angular momentum and spin-dependent interactions of the quarks are modified by the quark-gluon interaction of QCD. The full one-loop quark-antiquark interaction potential was obtained  \cite{gr1,bnt,gr2,npt,ptn} and employed in calculating the spectra and decays for the $c\bar{c}$ and $b\bar{b}$ systems in non-relativistic \cite{grr1} and relativistic model calculations \cite{grr2,grr3}. The spectra determined in these models proved to be very accurate for both $c\bar{c}$ and $b\bar{b}$. In the case of the $\Upsilon$ system, the addition of the one-loop corrections enabled an accurate prediction of the energies and splittings of the 1P, 2P and 1D levels \cite{grr1}.

More recently, the precise determination of leptonic and $E1$ decay widths, along with the discovery of several new states, some expected and some not, has led to a renewed interest in the potential model description of heavy quarkonia \cite{elq,elq1,bg,bg1,ls,DPF2004,panic2005}. We have revised and extended the approach of our earlier papers in order to investigate the newly measured states and decays as well as to discuss other questions of modelling interest. These include the scalar/vector mixture of the phenomenological confining potential, errors associated with the determination of the parameters appearing in the potential, and the accuracy of the perturbation expansion as determined by treating the Hamiltonian as ``complete'' in the variational calculation discussed below. 

In the next Section, we describe the potential model in some detail. This is followed, in Section \ref{III}, by an outline of our calculational approach. In Section \ref{IV}, we present our results for the charmonium and upsilon systems, and then give some conclusions in Section \ref{V}. Some calculational details are given in the appendix.

\section{SEMI-RELATIVISTIC MODEL \label{II}}

In our analysis, we use a semi-relativistic Hamiltonian of the form
\begin{eqnarray}
H&=&2\sqrt{\vec{p}^{\,2}+m^2}+Ar-\frac{4\al_S}{3r}\left[1-\frac{3\al_S}{2\pi}+\frac{\al_S}{6\pi}(33-2n_f) \left(\ln(\mu r)+\ga_E\right)\right] +V_L+V_S \\
&=&H_0+V_L+V_S\,,
\end{eqnarray}
where $\mu$ is the renormalization scale, $n_f$ is the effective number of light quark flavors and $\gamma_E$ is Euler's constant. $V_L$ contains the $v^2/c^2$ corrections to the linear confining potential
\begin{eqnarray} \label{VL}
V_{L}&=&-(1-f_V)\frac{A}{2m^2r}\vec{L}\Dot\vec{S}\nonumber \\
& & +f_V\left[\frac{A}{2m^2r}(1+\frac{8}{3}\vec{S_1}\Dot\vec{S_2})+ \frac{3A}{2m^2r} \vec{L}\Dot\vec{S}+\frac{A}{3m^2r}(3\vec{S_1}\Dot\hat{r} \vec{S_2}\Dot\hat{r}-\vec{S_1}\Dot\vec{S_2})\right],
\end{eqnarray}
where $A$ is the linear coupling strength. The first line in Eq.(3) is the contribution from scalar exchange while the second line is the contribution from vector exchange, with $f_V$ representing the fraction of vector exchange in the interaction. The short distance potential is \cite{gr1,bnt,gr2}
\begin{equation}\label{VS}
V_S=V_{HF}+V_{LS}+V_T+V_{SI},
\end{equation}
with
\begin{subequations}
\label{allpot}
\begin{eqnarray}
V_{HF}&=&\frac{32\pi\al_S\vec{S}_1\Dot\vec{S}_2}{9m^2}\left\{\left[1-\frac{\al_S}{12\pi}
(26+9\ln\,2)\right]\de(\vec{r})\right. \nonumber \\
& &\left.-\frac{\al_S}{24\pi^2}(33-2n_f)\nabla^2\left[\frac{\ln\,\mu r+\ga_E}{r}\right]+\frac{21\al_S}{16\pi^2}\nabla^2\left[\frac{\ln\, mr+\ga_E}{r}\right]\right\}\label{pota} \\
V_{LS}&=&\frac{2\al_S\vec{L}\Dot\vec{S}}{m^2r^3}\!\left\{1- \frac{\al_S}{6\pi}\left[\frac{11}{3}-(33-2n_f)\left(\ln\mu r+\ga_E-1\right)+12\left(\ln mr+\ga_E-1\right)\right]\right\}\label{potb}\\ [4pt]
V_{T\;}
&=&\frac{4\al_S(3\vec{S_1}\Dot\hat{r}\vec{S_2}\Dot\hat{r}-\vec{S_1}\Dot\vec{S_2})}
{3m^2r^3}\left\{1+\frac{\al_S}{6\pi}\left[8+(33-2n_f)\left(\ln\mu
r+\ga_E-\frac{4}{3}\right)\right.\right. \nonumber \\
& &\left.\left.-18\left(\ln mr+\ga_E-\frac{4}{3}\right)\right]\right\}\label{potc} \\
V_{SI}&=&\frac{4\pi\al_S}{3m^2}\left\{\left[1-\frac{\al_S}{2\pi}(1+\ln2)\right]
\de(\vec{r})-\frac{\al_S}{24\pi^2}(33-2n_f)\nabla^2\left[\frac{\ln\,\mu
r+\ga_E}{r}\right]-\frac{7\al_Sm}{6\pi r^2}\right\} \label{potd}\\\nonumber
\end{eqnarray}
\end{subequations}

We have chosen $H_0$ such that it contains the relativistic kinetic energy and the leading order spin-independent portions of the long-range confining potential and the one-loop QCD short-range potential. Note that, unlike all other terms in Eqs.\,(\ref{allpot}), the last term in Eq.\,(\ref{potd}) has a $1/m$ behavior.

\section{Calculational approach \label{III}}
The $c\bar{c}$ and $b\bar{b}$ mass spectra and corresponding wave functions are obtained using a variational approach. The wave functions were expanded as
\begin{equation}\label{wavefun}
\psi^m_{j\ell s}(\vec{r})=\sum_{k=0}^n C_k\left(\frac{r}{R}\right)^{k+\ell} \!e^{-r/R} \mathcal{Y}^m_{j\ell s}(\Omega)\,,
\end{equation}
where $\mathcal{Y}^m_{j\ell s}(\Omega)$ denotes the orbital-spin wave function for a specific total angular momentum $j$, orbital angular momentum $\ell$ and total spin $s$. The $C_k$'s are determined by minimizing
\begin{equation}E=\frac{\langle\psi\,|H\,|\psi\rangle} {\langle\psi\,|\psi\rangle}
\end{equation}
with respect to variations in these coefficients. This procedure results in a linear eigenvalue equation for the $C_k$'s and the energies (see Eq.\,(\ref{eigen})) and is equivalent to solving the Schr\"odinger equation. The wave functions corresponding to different eigenvalues are orthogonal and the $k^{\rm th}$ eigenvalue $\lambda_k$ is an upper bound on the exact energy $E_k$. For $n=14$, the lowest four eigenvalues for any $\ell$ are stable to one part in $10^6$.

We performed the calculations in two ways: (i) perturbatively, as implied above with $H_0$ as the unperturbed Hamiltonian and all other terms treated as first-order perturbations; (ii) nonperturbatively, with all terms included in the unperturbed Hamiltonian. The most significant effect of the different treatments is on the wave functions. Both approaches yield acceptable spectra, but the leptonic decay widths (which depend strongly on the behavior of the wave functions at the origin) turn out to be better in the perturbative treatment. On the other hand, the non-perturbative treatment provides a better description of certain $E1$ decays. The results of both approaches are shown below.

In either approach, we determine an optimal set of potential parameters  $\alpha=(\alpha_1,\alpha_2,\cdots,\alpha_n)$ by minimizing the $\chi^2$ function
\begin{equation}
\chi^2 = \sum_{i=1}^N\frac{\left(\mathcal{O}_{\rm exp\,i}-\mathcal{O}_{\rm th}(\alpha)_i\right)^2}{\sigma^2_i}\,,
\end{equation}
where the $\mathcal{O}_i$ denote the experimental and theoretical values of some quarkonium observable and the $\sigma_i$ are the associated errors. In this work, the $\mathcal{O}_{\rm exp\,i}$ consist of a subset of the measured onia masses and leptonic widths. For the masses, the $\sigma_i$ are taken to be the actual experimental error and a common intrinsic theoretical error added in quadrature. The latter error reflects the theory uncertainty associated with omitting corrections beyond one-loop and is estimated by requiring the $\chi^2$/degree of freedom to be approximately unity. Typically, this error is a few MeV. The minimization of $\chi^2$ with respect to  variations of the parameters $\alpha$ is accomplished using the search program STEPIT \cite{step}.

\section{RESULTS \label{IV}}

We summarize our results in the following tables. It should be noted that the parameter sets used in the charmonium and upsilon fits are independent of each other. The parameters resulting from our fits are given in Table \ref{params}. The errors on $A$, $\alpha_S$ and $m_q$ are those that change $\chi^2$ by approximately 1 when the parameters other than the one in question are allowed to vary.

\begin{table}[h]
\centering 
\begin{tabular}{ldddd}
\toprule
  &\multicolumn{1}{r}{\sb $c\bar{c}$\, Pert}\sw  &\multicolumn{1}{r}{\sb $c\bar{c}$\, Non-pert}\sw &
\multicolumn{1}{r}{\sb $b\bar{b}$\, Pert}\sw  &\multicolumn{1}{r}{
\sb $b\bar{b}$\, Non-pert}\sw \\
\hline
\sc$A$ (GeV$^2$) & 0.166_{-0.002}^{+0.002}   & 0.186_{-0.001}^{+0.003}   & 0.177_{-0.002}^{+0.006}    & 0.193_{-0.001}^{+0.004} \\
\hline
\sc $\al_S$      & 0.334_{-0.009}^{+0.009}   & 0.332_{-0.004}^{+0.003}   & 0.296_{-0.007}^{+0.004}    & 0.295_{-0.006}^{+0.002} \\
\hline
\sc $m_q$ (GeV)  & 1.51_{-0.08}^{+0.07}
    & 1.80_{-0.05}^{+0.03}    & 5.36_{-0.42}^{+0.87}     & 6.61_{-0.18}^{+0.35}  \\
\hline
\sc $\mu$ (GeV)  & 2.60    & 1.32    & 4.74     & 3.73  \\
\hline
\sc $f_V$        & 0.00    & 0.24    & 0.00     & 0.21  \\
\botrule
\end{tabular}
\caption{Fitted Parameters for the $c\bar{c}$ and
$b\bar{b}$ systems\label{params}}
\end{table}
As can be seen from Table \ref{params}, the value of the parameter $A$ characterizing the confinement strength is a smaller for the perturbative treatment compared to the non-perturbative treatment, but the variation is not great. The differences between perturbative $c\bar{c}$ and $b\bar{b}$ values for $A$ are rather small, as are the differences between the non-perturbative values. This may be an indication of some slight flavor dependence of the confining term. The values of $\alpha_S$ in the two treatments are very nearly identical for a given $q\bar{q}$ system and the difference between the $c\bar{c}$ and $b\bar{b}$ values is consistent with the fact that the scale in the upsilon system is higher. In each case, the value of the quark mass is smaller for the perturbative treatment compared to the non-perturbative treatment. The primary difference between the parameters obtained for the perturbative and non-perturbative approaches is in the values of the scale parameter $\mu$ and the fraction $f_V$ of vector exchange included in the confining potential. In both systems, the value of $f_V$ is zero in the perturbative case and of order 20\% in the non-perturbative case. Similarly, the renormalization scale is always larger for the perturbative treatment. We made no effort to restrict the value of $\mu$ while fitting other than to be sure that it didn't settle on a value that would suggest a change in the number of active flavors lighter than the quark flavor being considered.

\subsection{Charmonium}

The results for our determination of the charmonium levels are shown in Table \ref{charmspec}, where the $^*$ denotes the states used in the fit \cite{pdg}. We examined the effect of level mixing induced by the tensor interaction, and found that it has a very modest effect on the spectrum. However, as discussed below, $s$-$d$ mixing does have a significant effect on leptonic decay widths.

\begin{table}[h]\centering  
\begin{tabular}{lddd} \toprule
\multicolumn{1}{c}{\sc}$m_{c\bar{c}}$\,(MeV)  &\multicolumn{1}{c}{Pert}  &\multicolumn{1}{c}{ Non-pert}& \multicolumn{1}{c}{ Expt} \\
\hline
\sb$\eta_c(1S)^*$\mbox{\rule{12pt}{0pt}}   & 2980.3\sw\sw   & 2981.7  & 2980.4\pm 1.2    \\ 
\hline
\sb$\psi(1S)^*$     & 3097.36  & 3096.92    & 3096.916\pm 0.011  \\ 
\hline
\sb$\chi_{c\,0}(1P)^*$ & 3415.7   & 3415.2  & 3414.76\pm 0.35    \\
\hline
\sb$\chi_{c\,1}(1P)^*$ & 3508.2   & 3510.6  & 3510.66\pm 0.07    \\ 
\hline
\sb$\chi_{c\,2}(1P)^*$ & 3557.7   & 3556.2  & 3556.20\pm 0.09    \\
\hline
\sb$h_c(1P)$      & 3526.9   & 3523.7       & 3525.93\pm 0.27    \\
\hline
\sb$\eta_c(2S)$    & 3597.1   & 3619.2      & 3638.0\pm 4.0      \\
\hline
\sb$\psi(2S)^*$   & 3685.5   & 3686.1       & 3686.093\pm 0.034  \\
\hline
\sb$\psi(1D)$     & 3803.8   & 3789.4       & 3771.1\pm 2.4      \\
\hline
\sb$1^3D_2$     & 3823.8   & 3822.1         &                    \\
\hline
\sb$1^3D_3$     & 3831.1   & 3844.8         &                    \\
\hline
\sb$1^1D_2$     & 3823.6   & 3822.2         &                    \\
\hline
\sb$\chi_{c\,0}(2P)$ & 3843.7   & 3864.3    &                    \\
\hline 
\sb$\chi_{c\,1}(2P)$ & 3939.7   & 3950.0    &                    \\
\hline
\sb$\chi_{c\,2}(2P)$& 3993.7   & 3992.3   & 3929.\pm 5.4       \\
\hline
\sb$h_c(2P)$    & 3960.5   & 3963.2        &                     \\
\hline
\sb$1^3F_2$     & 4068.5   & 4049.9        &                     \\
\hline
\sb$1^3F_3$     & 4069.6   & 4069.0        &                     \\
\hline
\sb$1^3F_4$     & 4061.8   & 4084.3        &                     \\
\hline
\sb$1^1F_3$     & 4066.2   & 4066.9        &                     \\
\hline
\sb$\eta_c(3S)$ & 4014.0   & 4052.5        &                     \\
\hline
\sb$\psi(3S)  $ & 4094.9   & 4102.0        & 4039.\pm 1          \\
\hline
\sb$\psi(2D)^*$ & 4164.2   & 4159.2        & 4153.\pm 3          \\
\hline
\sb$2^3D_2$     & 4189.1   & 4195.8        &                     \\
\hline
\sb$2^3D_3$     & 4202.3   & 4218.9        &                     \\
\hline
\sb$2^1D_2$     & 4190.7   & 4196.9        &                     \\
\hline
\sb$\psi(4S)$   & 4433.3   & 4446.8        & 4421.\pm 4          \\
\hline
\sb$\psi(3D)$   & 4477.3   & 4478.9        &                     \\
\botrule
\end{tabular}
\caption{Perturbative and nonperturbative results for the $c\bar{c}$ spectrum are shown. The perturbative fit uses the indicated states and the leptonic widths of the $\psi(1S)$ and $\psi(2S)$. In the nonperturbative fit the $\eta_c(2S)$ and $\psi(1D)$ are included and no leptonic widths are used.}\label{charmspec}
\end{table}
In both the perturbative and non-perturbative treatments, the overall fit to the spectrum is quite good, with the $h_c(1P)$ mass well described and the $\eta_c(2S)$ mass somewhat low. Neither approach does very well in reproducing the $\psi(3S)$ level, although the $\psi(4S)$ level is accurately predicted. 

Recently, a number of states, which could be interpreted as above threshold $c\bar{c}$ states, have been observed in $B$ decays, at $e^+e^-$ colliders and at hadron colliders. Among these, the $X(3872)$ is the most firmly established \cite{Choi,Acosta}, with $m(X(3872))=3871.2\pm 0.5$\,MeV. An analysis of the quantum numbers of this state \cite{CDF} shows that the assignments $J^{PC}=1^{++},2^{-+}$ are the only ones capable of describing the data. If interpreted as charmonium states, the assignments are either $\chi_{c\,1}(2P)$ or $1^1D_2$. The former is more consistent with the masses in Table \ref{charmspec}, given the tendency of the potential model to predict a larger value for above threshold states. 

The Belle collaboration reported two states, the $X(3940)$, observed \cite{KAbe} in the recoil spectrum of $e^+e^-\to J/\psi\,X$, and the $Y(3940)$, observed \cite{SKChoi} in the decay $B\to K(\omega\,J/\psi)$. These appear to be different states \cite{KAbe}, with the former also seen to decay into $D^*\,\bar{D}$. Since the other states seen recoiling against the $J/\psi$ have $J=0$ \cite{KAbe}, it is tempting to associate $J=0$ with the $X(3940)$ \cite{ros}. The charmonium assignments could then be $\eta(3S)$, which would imply an unusually large $3S$ hyperfine splitting, or the $\chi_{c\,0}(2P)$, whose predicted mass tends to be too low. With the range of $\chi_{c\,J}(2P)$ masses in Table \ref{charmspec}, the $Y(3940)$ could be accommodated as a $2P$ state. 

The $BABAR$ collaboration reported a state $Y(4260)$ \cite{Aubert} observed in the initial state radiation (ISR) reaction $e^+e^-\to \gamma_{\rm\small IS}\,\pi^+\pi^-\,J/\psi$ and this state was confirmed by the CLEO collaboration \cite{Coan}. The ISR reaction assures $J^{PC}=1^{--}$, which, if the $Y(4260)$ is a conventional charmonium state, implies an assignment of $\psi(nS)$ or $\psi(nD)$. It has been suggested \cite{L-E} that the $Y(4260)$ rather than the $\psi(4415)$ be identified with the $\psi(4S)$, but this is at odds with both the absence of any $1^{--}$ level in the vicinity of this mass in most potential models and the absence of a large $\pi^+\pi^-\,J/\psi$ signal from the $\psi(4040)$ \cite{Coan}.

Included in Table \ref{charmspec} is the lowest multiplet of $f$-states. The center of mass of these states is essentially the same for both the perturbative and non-perturbative treatments, lying somewhat below the $\psi(3S)$. However, the pattern of splittings between the $^3F_J$ levels is quite different in the two cases.  

In computing the spectrum variationally, the wave functions are sampled in an average sense. The radiative and leptonic decay widths are of interest precisely because they are  sensitive to more detailed features of the wave functions, either matrix elements of the dipole operator in the case of $E1$ decays, or the behavior of the wave function at the origin in the case of leptonic decays. 

As is usual in potential model treatments \cite{grr3,elq1,bgs}, the radiative widths were calculated in the dipole approximation. We obtained the $E_1$ and $M_1$ matrix elements by using the variational radial wave functions to construct initial and final state wave functions with the appropriate angular dependence and explicitly performing the angular integration.  Our results are equivalent to the formulas
\begin{equation}
\Ga(n\,^3L_J\to n^\pr\,^3L^\pr_{J^\pr}+\ga)=\frac{4}{3}q^2\al\omega^3C_{fi}\left|\langle n^\pr\,
^3L^\pr_{J^\pr}\,|r|n\,^3L_J\rangle\right|^2\frac{E_f}{M_i}\,,
\end{equation}
for $E_1$ transitions, and
\begin{equation}
\Ga(n\,^{2s+1}L_J\to n^\pr\,^{2s^\pr+1}L_{J^\pr}+\ga)=\frac{4}{3}q^2\frac{\al}{m_q^2}\omega^3 \frac{2J^\pr+1}{2L+1}\de_{s\,s^\pr\pm 1}\left|\langle n^\pr\,^{2s^\pr+1}L_{J^\pr}\,| n\,^{2s+1}L_J\rangle \right|^2\frac{E_f}{M_i}\,,
\end{equation}
for $M_1$ transitions. Here, $\omega$ is the photon energy, $q$ is the quark charge in units of the proton charge, $E_f$ is the energy of the final quarkonium state, $M_i$ is the mass of the initial quarkonium state, $m_q$ is the quark mass and $C_{fi}$ contains the square of a $6j$ symbol that arises from the angular integration \cite{kros}. 

 The resulting $c\bar{c}$ radiative widths are shown in Table \ref{onegam} for both the perturbative and non-perturbative treatments. Comparing the two approaches, it 
\begin{table}[h]
\centering 
\begin{tabular}{lddd} \toprule
\multicolumn{1}{c}{\sc $\Gamma_{\ga}$\,(keV)}  &\multicolumn{1}{c}{Pert}  &\multicolumn{1}{c}{ Non-pert}& \multicolumn{1}{c}{ Expt}   \\
\hline
\sb$\psi(1S)\to\eta_c(1S)$\mbox{\rule{12pt}{0pt}}     & 2.7\sw\sw    & 1.8    & 1.21\pm 0.37   \\
\hline
\sb$\psi(2S)\to\eta_c(2S)$   & 1.2    & 0.4    & < 0.7          \\
\hline
\sb$\psi(2S)\to\eta_c(1S)$   & 0.0    & 0.45   & 0.88\pm 0.14   \\
\hline
\sb$\psi(2S)\to\chi_{c\,0}(1P)$   & 45.0   & 25.2   & 31.0\pm 1.8    \\
\hline
\sb$\psi(2S)\to\chi_{c\,1}(1P)$   & 40.9   & 29.1   & 29.3\pm 1.8    \\
\hline
\sb$\psi(2S)\to\chi_{c\,2}(1P)$   & 26.5   & 25.2   & 27.3\pm 1.7    \\
\hline
\sb$\eta_c(2S)\to h_c(1S)$   & 8.3    & 17.4   &                \\
\hline
\sb$\psi(3S)\to\chi_{c\,0}(2P)$   & 87.3   & 30.1   &                \\
\hline
\sb$\psi(3S)\to\chi_{c\,1}(2P)$   & 65.7   & 45.0   &                \\
\hline
\sb$\psi(3S)\to\chi_{c\,2}(2P)$   & 31.6   & 36.0   &                \\
\hline
\sb$\psi(3S)\to\chi_{c\,0}(1P)$   & 1.2    & 2.1    &                \\
\hline
\sb$\psi(3S)\to\chi_{c\,1}(1P)$   & 2.5    & 0.3    & < 880          \\
\hline
\sb$\psi(3S)\to\chi_{c\,2}(1P)$   & 3.3    & 2.4    & < 1360         \\
\hline
\sb $\chi_{c\,0}(1P)\to\psi(1S)$  & 142.2  & 139.3  & 135.\pm 15     \\
\hline
\sb $\chi_{c\,1}(1P)\to\psi(1S)$  & 287.0  & 293.7  & 317.\pm 25     \\ 
\hline
\sb $\chi_{c\,2}(1P)\to\psi(1S)$  & 390.6  & 384.1  & 417.\pm 32     \\
\hline
\sb $h_c(1P)\to\eta_c(1S)$   & 610.0  & 546.4  &                \\ 
\hline
\sb$\chi_{c\,0}(2P)\to\psi(2S)$   & 53.6   & 89.7   &                \\
\hline
\sb$\chi_{c\,1}(2P)\to\psi(2S)$   & 208.3  & 235.8  &                \\
\hline
\sb$\chi_{c\,2}(2P)\to\psi(2S)$   & 358.6  & 319.4  &                \\
\hline
\sb$\chi_{c\,0}(2P)\to\psi(1S)$   & 20.8   & 24.0   &                \\
\hline
\sb$\chi_{c\,1}(2P)\to\psi(1S)$   & 28.4   & 5.1    &                \\
\hline
\sb$\chi_{c\,2}(2P)\to\psi(1S)$   & 33.2   & 36.7   &                \\
\hline
\sb$\chi_{c\,0}(2P)\to\psi(1D)$   & 1.2    & 7.4    &                \\
\hline
\sb$\chi_{c\,1}(2P)\to\psi(1D)$   & 11.1   & 12.3   &                \\
\hline
\sb$\chi_{c\,2}(2P)\to\psi(1D)$   & 1.2    & 0.8    &                \\
\hline
\sb$\chi_{c\,1}(2P)\to 1^3D_2$    & 20.9   & 23.5   &                \\
\hline
\sb$\chi_{c\,2}(2P)\to 1^3D_2$    & 12.7   & 9.1    &                \\
\hline
\sb$\psi(1D)\to\chi_{c\,0}(1P)$   & 415.4  & 243.9  & 172.\pm 30^*    \\
\hline
\sb$\psi(1D)\to\chi_{c\,1}(1P)$   & 146.7  & 104.9  & 70.\pm 17^*    \\
\hline
\sb$\psi(1D)\to\chi_{c\,2}(1P)$   & 5.8    & 1.9    & < 21.^*      \\
\hline
\sb$1^3D_2\to\chi_{c\,1}(1P)$     & 317.3  & 256.7  &                \\
\hline
\sb$1^3D_2\to\chi_{c\,2}(1P)$     & 65.7   & 61.8   &                \\
\hline
\sb$1^3D_3\to\chi_{c\,2}(1P)$     & 62.7   & 39.5   &                \\
\hline
\sb$\psi(2D)\to\chi_{c\,0}(1P)$   & 8.9    & 23.3   &                \\
\hline
\sb$\psi(2D)\to\chi_{c\,1}(1P)$   & 4.7    & 0.02   & < 721.         \\
\hline
\sb$\psi(2D)\to\chi_{c\,2}(1P)$   & 0.26   & 0.23   & < 1340.        \\
\botrule
\end{tabular}
\caption{The the radiative decays of the charmonium system are shown. The $\psi(1D)\to\chi_J(1P)$ widths marked with a $^*$ are from \cite{CLEO}; see also \cite{ros}.}\label{onegam}
\end{table}
is noteworthy that there is a clear indication of a difference in behavior of the $p$ state wave functions. This is most easily seen in the results for the $\psi(2S)\to\chi_{c\,J}(1P)$ transition widths. In the perturbative case, the dipole matrix element is that same for every $J$, and the widths reflect the differences in phase space. On the other hand, the non-perturbative widths are very nearly equal, which means that the matrix elements for the various values of $J$ must differ. This difference is also reflected in predictions for the $\psi(1D)\to\chi_{c\,J}(1P)\,\gamma$ decays recently reported by the CLEO collaboration \cite{CLEO}. 

Unlike the radiative transitions, which are sensitive to matrix elements of the dipole operator, the leptonic widths are sensitive to the behavior of the wave function at the origin. Quite generally, the expression for the leptonic width of a $1^{--}$ state can be written
\begin{equation}
\Ga(1^{--}\to e\bar{e})=\frac{8\pi\al^2q^2}{3}\frac{|\langle 0|\vec{J}(0)|\psi_1^M\rangle|^2}{m_\psi^2}\,
\end{equation}
where $\langle 0|\vec{J}(0)|\psi_1^M\rangle$ is the matrix element of the quark current $i\bar{q}(0)\ga^\mu q(0)$ between the vacuum and the $1^{--}$ bound state. If $\Psi_{\al\be}(x)$ is the bound state two-body wave function, it is shown in Ref.\,\cite{hr} that 
\begin{equation}
\langle 0|J^\mu (0)|\psi_1^M\rangle = i\,{\rm Tr}[C^{-1}\ga^\mu\Psi(0)]\,,
\end{equation}
where $C$ is the charge conjugation matrix. This leads to the expressions \cite{hr}
\begin{equation} \label{rellep}
\Ga_{e\bar{e}}(\psi(nS))=\frac{4\al^2q^2}{m_{nS}^2}|I_{n0}|^2\qquad
\Ga_{e\bar{e}}(\psi(nD))=\frac{8\al^2q^2}{9m_{nD}^2}|I_{n2}|^2\,.
\end{equation} 
The integrals $I_{n0}$ and $I_{n2}$ are
\begin{equation}
I_{n0}=\sqrt{\frac{2}{\pi}}\int_0^\infty dpp^2\left(\frac{2}{3}+\frac{1}{3}\frac{m_q}{\sqrt{p^2+m_q^2}}\right)\phi_{n0}(p)\quad I_{n2}=\sqrt{\frac{2}{\pi}}\int_0^\infty dpp^2\left(1-\frac{m_q}{\sqrt{p^2+m_q^2}}\right)\phi_{n2}(p)\,.
\end{equation}
Here, $\phi_{n0}(p)$ and $\phi_{n2}(p)$ are the momentum space wave functions and the factors involving $\sqrt{p^2+m_q^2}$ represent relativistic corrections. To the leading order in $p^2/m_q^2$, these expressions reduce to the familiar forms 
\begin{equation}\label{nrlep}
\Gamma_{e\bar{e}}(\psi(nS))=\frac{4\alpha^2 q^2}{m^2_{nS}}|R_{n0}(0)|^2\quad
\Gamma_{e\bar{e}}(\psi(nD))=\frac{25\alpha^2 q^2}{2m^2_{nD} m_q^4}|R^{\,\prime\prime}_{n2}(0)|^2\,.
\end{equation}
For charmonium, we find that the relativistic effects on the $s$ states are rather small for both the perturbative and non-perturbative cases. The $d$ states, on the other hand, receive large corrections in both cases, with the non-perturbative case receiving the larger of the two. To be consistent, we have used the expressions in Eq.\,(\ref{rellep}) for all charmonium leptonic decay widths. For the $s$ states, we also include the QCD correction factor $(1 - 16\alpha_S/3\pi)$.  As mentioned above, we included mixing in our perturbative treatment and found little effect on the spectrum due to small mixing angles. However, even a small $s$-$d$ mixing angle has a noticeable effect on $\psi(nD)$ leptonic widths, and we included this mixing in the perturbative calculation. The results are shown in Table \ref{leptonic}.
\begin{table}[h]
\centering 
\begin{tabular}{lddd} \toprule
\multicolumn{1}{c}{\sc $\Gamma_{e\bar{e}}$\,(keV)}  &\multicolumn{1}{c}{Pert}  &\multicolumn{1}{c}{ Non-pert}& \multicolumn{1}{c}{\,\, Expt}   \\
\hline
\sb$\psi(1S)$\mbox{\rule{12pt}{0pt}} & 4.28\sw\sw  & 1.89  & 5.55\pm 0.14 \\
\hline
\sb$\psi(2S)$   & 2.25  & 1.04  & 2.48\pm 0.06 \\
\hline
\sb$\psi(3S)$   & 1.66  & 0.77  & 0.86\pm 0.07 \\
\hline
\sb$\psi(4S)$   & 1.33  & 0.65  & 0.58\pm 0.07 \\
\hline
\sb$\psi(1D)$   & 0.09  & 0.23  & 0.242\pm 0.030 \\
\hline
\sb$\psi(2D)$   & 0.16  & 0.45  & 0.83\pm 0.07 \\
\botrule
\end{tabular}
\caption{The leptonic widths of the $J=1^{--}$ states are shown.}\label{leptonic}
\end{table}
The message here is rather mixed in the sense the perturbative results for the $\psi(1S)$ and $\psi(2S)$ widths, which were included in the fit, are in reasonable agreement with the data, while the $\psi(3S)$ and $\psi(4S)$ widths are too large. Despite the inclusion of mixing, the $\psi(1D)$ and $\psi(2D)$ widths are small compared to the data. On the other hand, the non-perturbative treatment, which has no leptonic widths in the fit, does a reasonable job accounting for the $\psi(3S)$ and $\psi(4S)$ widths, but predicts $\psi(1S)$ and $\psi(2S)$ widths that are too small. With the inclusion of the relativistic correction, the non-perturbative $\psi(1D)$ and $\psi(2D)$ widths are of the right order of magnitude. 

\subsection{Upsilon}
The results for our determination of the upsilon levels are shown in Table \ref{bottomspec}, where, again,  the $^*$ denotes the states used in the fit. In this case, both approaches give very good fits to the spectrum. In the perturbative case, the determination of the parameters is quite robust in the sense that fits using fewer states than the eight indicated in Table \ref{bottomspec} yield parameters, masses and values of $\chi^2$ that are quite similar to the ones listed. For instance, using only the $\Up(1S),\Up(2S),\Up(3S)$ and $\Up(4S)$ states results in a fit in which the $1^3P_J$ multiplet is displaced upward by about $7$ MeV and the rest of the spectrum is well described. The nonperturbative fit includes the leptonic width of the $\Up(1S)$ and the main difference in the results of the two approaches occurs in the $\Up(nS)-\eta_b(nS)$ hyperfine splitting, which is always smaller in the nonperturbative treatment. In the $\Up$ system there are $f$ states occurring below $B\bar{B}$ threshold. The $1^1F_3,1^3F_J$ multiplet, in the vicinity of the $\Up(3S)$, is certainly below threshold and the $2^1F_3,2^3F_J$ multiplet is very near threshold.

Our results for the upsilon radiative transitions are shown in Table \ref{botgamma}. Due to a lack of knowledge of the widths of the $\chi_{b\,J}(1P)$ and $\chi_{b\,J}(2P)$ states, direct comparisons to experimental data are limited primarily to $\Up(nS)\to\chi_{b\,J}(n^\prime P)$ transitions. For these, both the perturbative and nonperturbative wave functions are able to account for the radiative widths quantitatively, including the $61\pm 23$\,eV width of the $\Up(3S)\to\chi_{b\,0}(1P)$ transition. Regarding the $\chi_{b\,J}(nP)\to\Up(n^\pr S))$ decays, one can compare  $\Gamma(\chi_{b\,J}(2P)\to\Up(2S))/\Gamma(\chi_{b\,J}(2P)\to\Up(1S))$ to the ratios of the measured branching ratios ($\Gamma_1/\Gamma_2$ in Ref. \cite{pdg}). These are appended to the bottom of Table \ref{botgamma}. Overall, both approaches give a good description of the data, with the perturbative results getting the nod in the case of $\Up(nS)\to\chi_{b\,J}(n^\prime P)$ transitions and the nonperturbative results being more consistent with the $\Ga_1/\Ga_2$ ratios. Given the current errors, though, the results from both approaches are certainly satisfactory.

The calculated leptonic widths for the $\Up(nS)$ states, including the QCD correction, are compared with the experimental data in Table \ref{botleptonic}. For the $\Up$ system, the relativistic corrections of Eq.\,(\ref{rellep}) are negligible. In the perturbative case, none of the leptonic width data were used in the fitting and the agreement is very good. For the nonperturbative case, where the $\Up(1S)$ leptonic width is used in the fit, the results are almost identical. Unlike the perturbative case, failure to include the $\Up(1S)$ leptonic width in the fit leads to an unsatisfactory description of all the $\Up(nS)$ leptonic widths. The calculated leptonic widths of the $\Up(1D)$ and $\Up(2D)$ are very small $\sim 0.02$ keV in both perturbative and nonperturbative treatments.
\begin{table}[h]\centering  
\begin{tabular}{lddd} \toprule
\multicolumn{1}{c}{\sc}$m_{b\bar{b}}$\,(MeV)  &\multicolumn{1}{c}{Pert}  & \multicolumn{1}{c}{ Non-pert}& \multicolumn{1}{c}{ Expt} \\
\hline
\sb$\eta_b(1S)$\mbox{\rule{12pt}{0pt}}   & 9413.70\sw\sw   & 9421.02  &    \\ 
\hline
\sb$\Up(1S)^*$         & 9460.69   & 9460.28  & 9460.30\pm 0.26  \\ 
\hline
\sb$\chi_{b\,0}(1P)^*$ & 9861.12   & 9860.43  & 9859.44\pm 0.52  \\
\hline
\sb$\chi_{b\,1}(1P)^*$ & 9891.33   & 9892.83  & 9892.78\pm 0.40  \\ 
\hline
\sb$\chi_{b\,2}(1P)^*$ & 9911.79   & 9910.13  & 9912.21\pm 0.40  \\
\hline
\sb$h_b(1P)$      & 9899.99   & 9899.94    &   \\
\hline
\sb$\eta_b(2S)$   & 9998.69   & 10003.6    &   \\
\hline
\sb$\Up(2S)^*$    & 10022.5   & 10023.5    & 10023.26\pm 0.31 \\
\hline
\sb$\Up(1D)$      & 10149.5   & 10148.8    &    \\
\hline
\sb$1^3D_2$       & 10157.1   & 10157.0    & 10161.1\pm 1.7    \\
\hline
\sb$1^3D_3$       & 10162.9   & 10164.1    &    \\
\hline
\sb$1^1D_2$       & 10158.4   & 10158.3    &    \\
\hline
\sb$\chi_{b\,0}(2P)^*$ & 10230.5   & 10231.4    & 10232.5\pm 0.6  \\
\hline 
\sb$\chi_{b\,1}(2P)^*$ & 10255.0   & 10257.6    & 10255.46\pm 0.55\\
\hline
\sb$\chi_{b\,2}(2P)^*$ & 10271.5   & 10271.1    & 10268.65\pm 0.55\\
\hline
\sb$h_b(2P)$    & 10262.0   & 10263.1    &                  \\
\hline
\sb$1^3F_2$     & 10353.0   & 10351.0    &                 \\
\hline
\sb$1^3F_3$     & 10355.8   & 10355.6    &                 \\
\hline
\sb$1^3F_4$     & 10357.5   & 10359.7    &                 \\
\hline
\sb$1^1F_3$     & 10355.9   & 10355.9    &                 \\
\hline
\sb$\eta_b(3S)$ & 10344.8   & 10350.4    &                  \\
\hline
\sb$\Up(3S)$    & 10363.6   & 10365.6    & 10355.2\pm 0.5   \\
\hline
\sb$\Up(2D)$    & 10443.1   & 10443.7    &                  \\
\hline
\sb$2^3D_2$     & 10450.3   & 10451.2    &                  \\
\hline
\sb$2^3D_3$     & 10455.9   & 10457.5    &                  \\
\hline
\sb$2^1D_2$     & 10451.6   & 10452.4    &                  \\
\hline
\sb$2^3F_2$     & 10610.0   & 10609.0    &                 \\
\hline
\sb$2^3F_3$     & 10613.0   & 10613.4    &                 \\
\hline
\sb$2^3F_4$     & 10615.0   & 10617.3    &                 \\
\hline
\sb$2^1F_3$     & 10613.2   & 10613.7    &                 \\
\hline
\sb$\eta_b(4S)$ & 10622.8   & 10631.5    &                  \\
\hline
\sb$\Up(4S)$    & 10643.0   & 10643.4    & 10579.4\pm 1.2   \\
\botrule
\end{tabular}
\caption{Perturbative and nonperturbative results for the $b\bar{b}$ spectrum are shown. The perturbative fit uses the indicated states .}\label{bottomspec}
\end{table}

\begin{table}[h]
\centering 
\begin{tabular}{lddd} \toprule
\multicolumn{1}{c}{\sc $\Gamma_{\ga}$\,(keV)}  &\multicolumn{1}{c}{Pert}  &\multicolumn{1}{c}{ Non-pert}& \multicolumn{1}{c}{ Expt}   \\
\hline
\sb$\Up(1S)\to\eta_b(1S)$\mbox{\rule{12pt}{0pt}}     & 0.004\sw\sw    & 0.001    &    \\
\hline
\sb$\Up(2S)\to\eta_b(2S)$        & 0.0005 & 0.0002 &                 \\
\hline
\sb$\Up(2S)\to\eta_b(1S)$        & 0.0    & 0.005  & <0.02           \\
\hline
\sb$\Up(2S)\to\chi_{b\,0}(1P)$   & 1.15   & 0.74   & 1.22\pm 0.16    \\
\hline
\sb$\Up(2S)\to\chi_{b\,1}(1P)$   & 1.87   & 1.40   & 2.21\pm 0.22    \\
\hline
\sb$\Up(2S)\to\chi_{b\,2}(1P)$   & 1.88   & 1.67   & 2.29\pm 0.22    \\
\hline
\sb$\eta_b(2S)\to h_b(1P)$       & 4.17   & 20.4   &                 \\
\hline
\sb$\Up(3S)\to\chi_{b\,0}(2P)$   & 1.67   & 1.07   & 1.20\pm 0.16    \\
\hline
\sb$\Up(3S)\to\chi_{b\,1}(2P)$   & 2.74   & 2.05   & 2.56\pm 0.34    \\
\hline
\sb$\Up(3S)\to\chi_{b\,2}(2P)$   & 2.80   & 2.51   & 2.66\pm 0.41    \\
\hline
\sb$\Up(3S)\to\chi_{b\,0}(1P)$   & 0.03   & 0.03   & 0.061\pm 0.023  \\
\hline
\sb$\Up(3S)\to\chi_{b\,1}(1P)$   & 0.09   & 0.003  &                \\
\hline
\sb$\Up(3S)\to\chi_{b\,2}(1P)$   & 0.13   & 0.11   &                \\
\hline
\sb $\chi_{b\,0}(1P)\to\Up(1S)$  & 22.1   & 19.6   &                \\
\hline
\sb $\chi_{b\,1}(1P)\to\Up(1S)$  & 27.3   & 23.9   &                \\ 
\hline
\sb $\chi_{b\,2}(1P)\to\Up(1S)$  & 31.2   & 26.3   &                \\
\hline
\sb $h_b(1P)\to\eta_b(1S)$       & 37.9   & 4.61   &                \\ 
\hline
\sb$\chi_{b\,0}(2P)\to\Up(2S)$   & 9.90   & 9.91   &                \\
\hline
\sb$\chi_{b\,1}(2P)\to\Up(2S)$   & 13.7   & 12.4   &                \\
\hline
\sb$\chi_{b\,2}(2P)\to\Up(2S)$   & 16.8   & 13.5   &                \\
\hline
\sb$\chi_{b\,0}(2P)\to\Up(1S)$   & 6.69   & 1.83   &                \\
\hline
\sb$\chi_{b\,1}(2P)\to\Up(1S)$   & 7.31   & 4.81   &                \\
\hline
\sb$\chi_{b\,2}(2P)\to\Up(1S)$   & 7.74   & 6.86   &                \\
\hline
\sb$\chi_{b\,0}(2P)\to\Up(1D)$   & 1.13   & 1.05    &                \\
\hline
\sb$\chi_{b\,1}(2P)\to\Up(1D)$   & 0.62   & 0.52   &                \\
\hline
\sb$\chi_{b\,2}(2P)\to\Up(1D)$   & 0.04   & 0.03   &                \\
\hline
\sb$\chi_{b\,1}(2P)\to 1^3D_2$   & 1.48   & 1.31   &                \\
\hline
\sb$\chi_{b\,2}(2P)\to 1^3D_2$   & 0.47   & 0.35    &               \\
\hline
\sb$\Up(1D)\to\chi_{b\,0}(1P)$   & 18.1   & 12.5   &                \\
\hline
\sb$\Up(1D)\to\chi_{b\,1}(1P)$   & 9.82   & 7.59   &                \\
\hline
\sb$\Up(1D)\to\chi_{b\,2}(1P)$   & 0.51   & 0.44   &                \\
\hline
\sb$1^3D_2\to\chi_{b\,1}(1P)$    & 19.3   & 14.9   &                \\
\hline
\sb$1^3D_2\to\chi_{b\,2}(1P)$    & 5.07   & 4.35   &                \\
\hline
\sb$1^3D_3\to\chi_{b\,2}(1P)$    & 21.7   & 18.8   &                \\
\botrule
\multicolumn{1}{c}{\sc $\Ga_1/\Ga_2$}  &\multicolumn{1}{c}{Pert}  &\multicolumn{1}{c}{ Non-pert}& \multicolumn{1}{c}{ Expt}   \\
\hline
\sb$\Ga_1(\chi_{b\,0})/\Ga_2(\chi_{b\,0})$   & 1.48 & 5.42 & 5.11\pm 4.14\\
\hline
\sb$\Ga_1(\chi_{b\,1})/\Ga_2(\chi_{b\,1})$   & 1.87 & 2.58 & 2.47\pm 0.60\\
\hline
\sb$\Ga_1(\chi_{b\,2})/\Ga_2(\chi_{b\,2})$   & 2.17 & 1.97 & 2.28\pm 0.47\\
\botrule
\end{tabular}
\caption{The radiative decays of the upsilon system are shown.}\label{botgamma}
\end{table}
\begin{table}[h]
\centering 
\begin{tabular}{lddd} \toprule
\multicolumn{1}{c}{\sc $\Gamma_{e\bar{e}}$\,(keV)}  &\multicolumn{1}{c}{Pert}  &\multicolumn{1}{c}{ Non-pert}& \multicolumn{1}{c}{\,\, Expt}   \\
\hline
\sb$\Up(1S)$\mbox{\rule{12pt}{0pt}} & 1.33\sw\sw  & 1.33  & 1.340\pm 0.018 \\
\hline
\sb$\Up(2S)$   & 0.61  & 0.62  & 0.612\pm 0.011 \\
\hline
\sb$\Up(3S)$   & 0.46  & 0.45  & 0.443\pm 0.008 \\
\hline
\sb$\Up(4S)$   & 0.35  & 0.30  & 0.272\pm 0.029 \\
\botrule
\end{tabular}
\caption{The leptonic widths of the $\Up(nS)$ states are shown.}\label{botleptonic}
\end{table}

\section{CONCLUSIONS \label{V}}

We have shown that a potential model consisting of the relativistic kinetic energy, a linear long-range confining potential together with its $v^2/c^2$ relativistic corrections, and the full $v^2/c^2$ plus one-loop QCD corrected short distance potential is capable of providing extremely good fits to the spectra of the $c\bar{c}$ and $b\bar{b}$ heavy quarkonium systems. Interestingly enough, the the results are about as good with the spin-dependent and non-leading order spin-independent terms of the potential treated either as a perturbation or as part of the unperturbed Hamiltonian. We find that for both the charmonium and upsilon systems, the perturbative treatment requires the long-range potential to be entirely due to scalar exchange, while the non-perturbative treatment requires the long-range potential to be about one-fifth vector exchange. 

The photon and leptonic widths obtained from the variational wave functions are, for the most part, in very good agreement with the available data. An interesting difference between the two treatments is that in the non-perturbative treatment every state has its own wave function, while in the perturbative treatment all states in the same angular momentum multiplet have the same wave function. It is this feature that is responsible nonperturbative treatment's somewhat better description of the radiative widths. In the charmonium system, the leptonic widths are better described by the perturbative wave functions, while in the upsilon system, the nonperturbative description is just as good as the perturbative description provided one leptonic width is included in the fit. In both the charmonium and upsilon systems, inclusion of the QCD correction to the $s$ state leptonic widths is essential to obtain agreement with experiment.

Although in some respects the perturbative treatment, and hence pure scalar exchange for the long range potential, yields a somewhat better description of the charmonium system, the non-perturbative approach, with mixed scalar and vector exchange, is also viable. In the upsilon system, both approaches give very good descriptions of the available data and, given the rich spectrum below the continuum threshold, it should be possible to decide which is preferable with additional data such as the $\Up(nS)-\eta_b(nS)$ hyperfine splittings.

\begin{acknowledgments}
We would like to thank Michael Saelim for developing an important piece of code used in this analysis. This research was supported in part by the National Science Foundation under Grants PHY-0244789 and PHY-0555544.
\end{acknowledgments}

\appendix
\section{Some calculational details \label{VI}}

The choice of the variational wave function, Eq.\,(\ref{wavefun}) results in an equation for the $C_k$'s and the energies of the form
\begin{equation} \label{eigen}
\sum_{k=0}^n\langle i\,|H|k\rangle C_k=\lambda\sum_{j=0}^n\langle i\,|N|k\rangle C_k\,,
\end{equation}
where for a state with fixed $j$,$\ell$ and $s$
\begin{eqnarray}
\langle i\,|H|k\rangle &=& \int d^3r\left(\frac{r}{R}\right)^{i+\ell}e^{-r/R} \left(\mathcal{Y}^m_{j\ell s}(\Omega)\right)^\dagger H \left(\frac{r}{R}\right)^{k+\ell}e^{-r/R} \mathcal{Y}^m_{j\ell s}(\Omega)\,, \\
\langle i\,|N|k\rangle &=& \int d^3r\left(\frac{r}{R}\right)^{i+k+2\ell} e^{-2r/R}\left(\mathcal{Y}^m_{j\ell s}(\Omega)\right)^\dagger\mathcal{Y}^m_{j\ell s}(\Omega) =\frac{R^3}{2^{i+k+2\ell+3}}\Gamma(i+k+2\ell+3)\,. 
\end{eqnarray}
The matrix elements $\langle i\,|V|k\rangle$ of the potential, Eqs.\,(\ref{VL}) and (\ref{VS}), can be evaluated analytically in coordinate space. The contribution of the kinetic energy, $2\sqrt{\vec{p}^{\,2}+m^2}$, is evaluated numerically in momentum space, where the expansion of the wave function is
\begin{equation} \label{momwavefun}
\phi_{j\ell s}^m(\vec{p})=(-i)^\ell R^3\sum_{k=0}^nC_k \frac{\left(\cos\theta\right)^{k+\ell+2}}{\tan\theta}\sqrt{\sin\theta}\, \Gamma(k+2\ell+3)P^{-\ell-1/2}_{k+\ell+3/2}(\cos\theta) \mathcal{Y}_{j\ell s}^m(\Omega)\,.
\end{equation}
Here, $P^{-\ell-1/2}_{k+\ell+3/2}(\cos\theta)$ is the associated Legendre function and $\theta$ is related to the magnitude of $\vec{p}$ as
\begin{equation}
\tan\theta=pR\,.
\end{equation}

Given the matrix elements  $\langle i\,|H|k\rangle$ and $\langle i\,|N|k\rangle$,  eigenvalues and eigenvectors for the linear system Eq.\,(\ref{eigen}) can be obtained using any standard package \cite{IMSL}. We typically use 14 terms in the expansion Eq.\,(\ref{wavefun}).

Calculation of the matrix elements of potential using the variational wave function Eq.\,(\ref{wavefun}) is relatively straight forward. Perhaps the only integrals requiring special attention are those involving
\begin{equation}\label{delsq}
\int_0^\infty\!drr^2\nabla^2\left[\frac{\ln\,\mu r+\ga_E}{r}\right]F(r)\,,
\end{equation}
where $F(r)$ denotes a product of the radial wave functions, which has the form  $(r/R)^{L+L^\pr}e^{-2r/R}$. If $L+L^\pr \neq 0$, then, effectively, 
\begin{equation}
\nabla^2\left[\frac{\ln\,\mu r+\ga_E}{r}\right]\to -\frac{1}{r^3}\,,
\end{equation}
and the integral in Eq.\,(\ref{delsq}) reduces to a Gamma function. When $L+L^\pr=0$, we have
\begin{equation}
\frac{1}{4\pi}\int_0^\infty\!drr^2\nabla^2\left[\frac{\ln\,\mu r+\ga_E}{r}\right]e^{-2r/R} = 1 - \ln\left(\frac{\mu R}{2}\right)\,.
\end{equation}
When calculating the matrix elements of the $\delta(\vec{r})$ terms in Eqs.\,(\ref{pota}) and (\ref{potd}), we `soften' their singularity by adopting the quasistatic approximation of Ref.\cite{gjrs}, which leads to the replacement
\begin{equation}
\de(\vec{r})\to\frac{m^2}{\pi r}e^{-2mr}\,
\end{equation}
where $m$ is the quark mass. This softening helps the stability of the eigenvalue calculation, particularly in the non-perturbative approach.

\end{document}